\documentclass[sigconf, preprint]{acmart}

\usepackage{xcolor}
\usepackage{mdframed}
\usepackage{float}
\usepackage{booktabs}
\usepackage{caption}
\usepackage[fixed]{fontawesome5}
\faStyle{regular}
\usepackage{enumitem}
\usepackage{graphicx}
\usepackage{subcaption}
\AtBeginDocument{%
  \providecommand\BibTeX{{%
    \normalfont B\kern-0.5em{\scshape i\kern-0.25em b}\kern-0.8em\TeX}}}

\renewcommand\footnotetextcopyrightpermission[1]{} 
\acmConference[Preprint]{}{February 2024}{Toronto}

\begin{document}

\title[Supporting Self-Reflection at Scale with Large Language Models]{Supporting Self-Reflection at Scale with Large Language Models: Insights from Randomized Field Experiments in Classrooms}

\author{Harsh Kumar}
\email{harsh@cs.toronto.edu}
\orcid{0000-0003-2878-3986}
\affiliation{%
  \institution{University of Toronto}
  \country{Canada}
}

\author{Ruiwei Xiao}
\email{ruiweix@andrew.cmu.edu}
\affiliation{%
 \institution{Carnegie Mellon University}
  \country{United States}
  }
  
\author{Benjamin Lawson}
\email{benjamin.lawson@mail.utoronto.ca}
\affiliation{%
  \institution{University of Toronto}
  \country{Canada}
}

\author{Ilya Musabirov}
\email{imusabirov@cs.toronto.edu}
\affiliation{%
  \institution{University of Toronto}
  \country{Canada}
}

\author{Jiakai Shi}
\email{jiakai.shi@mail.utoronto.ca}
\affiliation{%
  \institution{University of Toronto}
  \country{Canada}
}

\author{Xinyuan Wang}
\email{xyben.wang@mail.utoronto.ca}
\affiliation{%
  \institution{University of Toronto}
  \country{Canada}
}

\author{Huayin Luo}
\email{huayin.luo@mail.utoronto.ca}
\affiliation{%
  \institution{University of Toronto}
  \country{Canada}
}

\author{Joseph Jay Williams}
\email{williams@cs.toronto.edu}
\affiliation{%
 \institution{University of Toronto}
  \country{Canada}
  }

\author{Anna Rafferty}
\email{arafferty@carleton.edu}
\affiliation{%
 \institution{Carleton College}
  \country{United States}
  }
  
\author{John Stamper}
\email{jstamper@cmu.edu}
\affiliation{%
 \institution{Carnegie Mellon University}
  \country{United States}
  }
  
\author{Michael Liut}
\email{michael.liut@utoronto.ca}
\orcid{0000-0003-2965-5302}
\affiliation{%
  \institution{University of Toronto Mississauga}
  \country{Canada}
}
\renewcommand{\shortauthors}{Kumar, et al.}

\begin{abstract}
Self-reflection on learning experiences constitutes a fundamental cognitive process, essential for the consolidation of knowledge and the enhancement of learning efficacy. However, traditional methods to facilitate reflection often face challenges in personalization, immediacy of feedback, engagement, and scalability. Integration of Large Language Models (LLMs) into the reflection process could mitigate these limitations. In this paper, we conducted two randomized field experiments in undergraduate computer science courses to investigate the potential of LLMs to help students engage in post-lesson reflection. In the first experiment (N=145), students completed a take-home assignment with the support of an LLM assistant; half of these students were then provided access to an LLM designed to facilitate self-reflection. The results indicated that the students assigned to LLM-guided reflection reported increased self-confidence and performed better on a subsequent exam two weeks later than their peers in the control condition. In the second experiment (N=112), we evaluated the impact of LLM-guided self-reflection against other scalable reflection methods, such as questionnaire-based activities and review of key lecture slides, after assignment. Our findings suggest that the students in the questionnaire and LLM-based reflection groups performed equally well and better than those who were only exposed to lecture slides, according to their scores on a proctored exam two weeks later on the same subject matter. These results underscore the utility of LLM-guided reflection and questionnaire-based activities in improving learning outcomes. Our work highlights that focusing solely on the accuracy of LLMs can overlook their potential to enhance metacognitive skills through practices such as self-reflection. We discuss the implications of our research for the edtech community, highlighting the potential of LLMs to enhance learning experiences through personalized, engaging, and scalable reflection practices.
\end{abstract}





\maketitle

\section{Introduction}
\label{section:intro}

Self-reflection on learning experiences is a critical metacognitive skill of the educational process \cite{wilson2016anderson}, which offers substantial benefits for both knowledge retention and self-efficacy \cite{bannert_assessment_2008, bielaczyc_training_1995, denton_effects_2023}. For example, when students engage in self-reflection after completing an assignment, this practice can significantly enhance their performance on subsequent exams \cite{davis_scaffolding_2000, zohar_explicit_2008}. Such reflective activities encourage students to revisit and consolidate what they have learned, fostering a deeper understanding and facilitating the application of knowledge in different contexts \cite{chi_eliciting_1994, iordanou_supporting_2022, lin_promoting_2012}.

Researchers have documented the extensive benefits of reflection practices, facilitated through various prompts and methodologies \cite{brown_visualizing_2010, govaerts_student_2012, johnston_amplifying_2005, leijen_streaming_2009, nelson_push-me_2008, palinscar_reciprocal_1984, reznitskaya_examining_2012, tseng_design_2013, webb_promoting_2013, shi_enhancing_2019, bentvelzen2022revisiting}. Specifically, the literature demonstrates that encouraging students to engage in reflection activities after interacting with the course materials can lead to gains in knowledge and skill acquisition \cite{chen2017strategic, menekse2022different}. This is achieved by promoting connections between disparate concepts within academic material \cite{davis_scaffolding_2000, bielaczyc_training_1995, denton_effects_2023}, encouraging the elaboration of learned ideas \cite{davis_scaffolding_2000}, and facilitating error detection and correction \cite{chi_eliciting_1994, davis_scaffolding_2000}. The underlying mechanism for these benefits involves structured metacognition examinations, coupled with opportunities for feedback from instructors or peers \cite{iordanou_theory_2016, felton_development_2004, zohar_explicit_2008}. This process allows students to better assess their understanding of the topic and explore effective strategies to apply their knowledge.

However, the logistics of most educational programs necessarily impose limitations on the amount, quality, and scalability of opportunities available to students to engage in these types of reflective activities. Class time where this would be possible, as well as that specifically allotted to review and question answering, is constrained by the amount of time required to teach the relevant course material in class and the availability of the instructor or teaching assistants outside of class time. Coursework assigned to be done outside of class time focusing on reflection will be subject to delayed feedback as well as little potential for personalization. Even when students directly interact with course staff at times like office hours, the amount of time available for students to reflect on content and receive immediate feedback from the course staff is limited.

The advent of Large Language Models (LLMs), such as GPT-4, heralds a new era in educational support tools, particularly in the context of online and hybrid learning environments that have gained prominence following the COVID-19 pandemic. LLMs, characterized by their ability to generate human-like text based on vast amounts of data, excel at engaging users in fluent multi-turn conversations in real time \cite{shanahan2023role}. This capability makes them particularly suited to facilitate reflective practices and personalized learning experiences. Researchers have shown that LLMs can function effectively as adaptive tutors, providing instruction and feedback tailored to the individual needs and progress of students. By simulating one-on-one tutoring sessions, these models can guide students through course material with a level of personalization previously difficult to achieve on a scale, potentially significantly enhancing learning outcomes \cite{baidoo-anu_education_2023, chen_personalized_2020}. Early evidence suggests that access to these personalized conversational AI tools could not only support knowledge acquisition \cite{kumar2023math} but also facilitate a growth mindset in learners \cite{handa2023mistakes}, making them a valuable asset in the educational toolkit.

Despite the considerable potential of LLMs to improve educational practices at scale, the novelty of this technology means that research on its efficacy, particularly in promoting specific educational strategies such as reflection techniques, remains sparse \cite{murugesan2023rise}. 
The lack of empirical evidence leaves open questions regarding the most effective ways to integrate LLMs into educational practices. Moreover, while LLM-based dialogue agents present a frontier of educational technology, they fundamentally differ from human tutors in several aspects. Unlike humans, whose language skills are deeply intertwined with cognitive capacities developed through embodied interactions within a community, LLMs operate as disembodied neural networks \cite{shanahan2023role}. Trained on extensive corpora of human-generated text, their primary function is to predict subsequent words in a sequence, lacking the nuanced understanding and experiences that come from living within a physical world. This distinction underscores the need to carefully consider how LLMs are deployed for educational purposes, particularly to foster effective collaborative learning environments. As we are on the brink of an era where collaborative learning with AI becomes a tangible reality, it is essential to gather more empirical evidence on how such collaborations with AI can be most effectively realized.




In this paper, we explore the efficacy of LLM-guided self-reflection in educational settings, specifically examining its impact on student learning outcomes and self-efficacy. Our investigation is structured around two studies conducted within classroom environments, comparing LLM-guided self-reflection to both a control group with no reflection intervention and other scalable reflection-promoting methods. The outcome measures include the performance of the students in subsequent proctored tests and self-reported metrics, such as confidence in the subject matter. The research questions are reformulated as follows.

\begin{itemize}
    \item [\textbf{RQ1}] What is the impact of self-reflection guided by LLM on learning outcomes and self-efficacy compared to no reflection intervention?
    \item [\textbf{RQ2}] Compared to conventional scalable reflection methods (such as a questionnaire-based self-reflection, or review of key lecture slides), what impact does LLM-guided self-reflection have on learning outcomes and self-confidence?
\end{itemize}


To address RQ1, we conducted a randomized field experiment ($N=145$) in an undergraduate computer science course on databases. All students completed a take-home assignment with the help of a GPT-3-based assistant, as detailed in Section \ref{section: exp1}. Subsequently, the students were randomly divided into two groups: One group was granted access to an LLM specifically designed to promote self-reflection on the assignment's material and concepts, while the other group did not engage in any reflection. Students who participated in the LLM-guided reflection process reported a notable increase in self-confidence. 
An accompanying qualitative analysis of the conversation logs revealed that the LLM provided positive feedback and affirmations, expanded on the students' answers, offered guidance, and encouraged reflection.

For RQ2, we conducted another randomized field experiment ($N=112$) in a first-year undergraduate computer programming course. This experiment was designed to compare the effectiveness of different self-reflection methodologies in helping students prepare for exams. As a component of the documentation for the take-home assignment, students were provided with a link that offered the opportunity to optionally reflect on the concepts they had learned. Upon accessing the link, students were randomly assigned to one of three distinct groups: the first group encountered a questionnaire-based reflection activity designed to stimulate reflection on the assignment's concepts; the second group received access to an LLM tailored to facilitate in-depth self-reflection on the learned material; and the third group was presented with curated snippets from the most important slides from the lecture related to the assignment topic. This setup allowed for a direct comparison of the impact of different reflection methods with a prevalent method students use for learning and exam preparation. We found that students assigned to the LLM-based reflection condition performed on a level similar to that of students in the static questionnaire condition and apparently better than students who were assigned to revise the curated set of slides.
%

Our research highlights that focusing solely on the accuracy of LLMs can overlook their potential to enhance metacognitive skills through practices such as self-reflection. LLMs provide a way to ``talk things out'' that may improve self-efficacy, and where the emphasis on reflection might decrease the need for full content accuracy.

\section{Related Work}
\label{section:background}
\subsection{Role of Reflection in Education}
Many have previously argued for and demonstrated the value of reflection in education \cite{bannert_assessment_2008, bielaczyc_training_1995, brown_visualizing_2010, chi_eliciting_1994, davis_scaffolding_2000, denton_effects_2023, govaerts_student_2012, iordanou_supporting_2022, johnston_amplifying_2005, leijen_streaming_2009, lin_promoting_2012, nelson_push-me_2008, palinscar_reciprocal_1984, reznitskaya_examining_2012, tseng_design_2013, webb_promoting_2013, zohar_explicit_2008, shi_enhancing_2019}. Researchers such as Lamberty and Koloder \cite{lamberty_camera_2005} who had students speak out loud to a camera when describing their work noted the reflective nature of their comments as well as the benefits to their mathematical understanding. While there has been limited experimental literature, several papers conducted behavioral experiments on the impact of reflection on student outcomes. 
Denton and Ellis examined techniques to improve comprehension of social studies in 7th grade students, wherein students were either assigned to one of two reflection-based conditions whereby they would answer either open or closed reflection prompts for which they would later receive feedback in the last 5 minutes of class-time, or a control group where they simply received 5 extra minutes of instruction \cite{denton_effects_2023}. Students in the reflection prompt groups significantly outperformed those in the control condition. 

Similar to reflection, researchers have explored how to support students in self-regulated learning and developing metacognitive skills \cite{azevedo2005scaffolding, azevedo2009metatutor, antonietti2015metacognition}. Our work contributes to a wider body of work that has focused on how to support students in their learning strategies \cite{johnson2018pedagogical}, and research that has looked at different levels of support needed and how that level of support impacts learners in both current and future interactions.

\begin{figure*}
    \includegraphics[width=\textwidth]{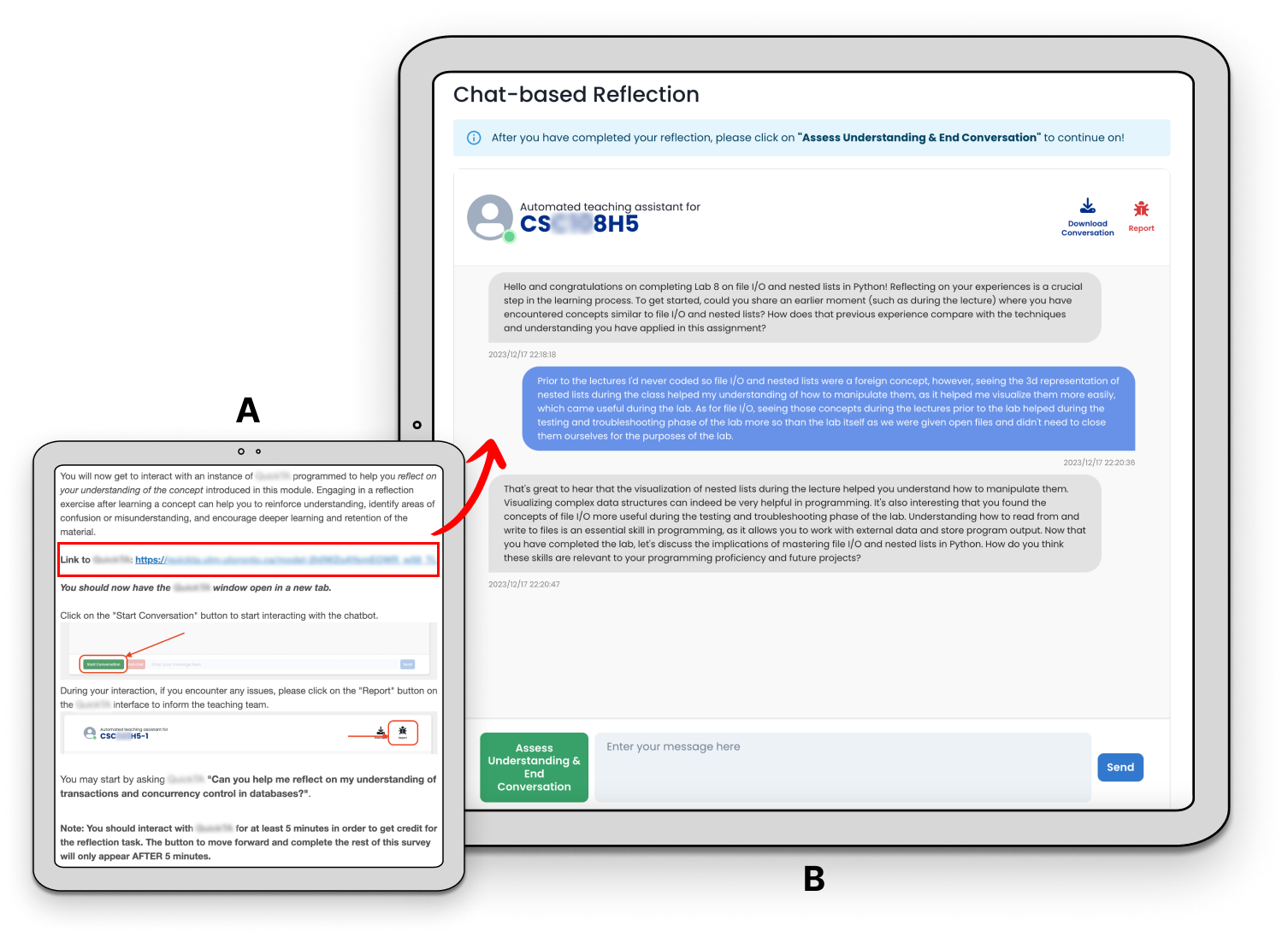}
    \caption{Stimuli for Study-1. A) Half of the students got link to the reflection bot after completing their assignment. B) Example chat window for reflection.}
    \label{fig:quickta_combined}
\end{figure*}

\subsection{LLMs in Classroom}
 LLMs and conversational AI systems have been acknowledged as game changers for science, and many have theorized about their potential applications in the field of education \cite{van_dis_chatgpt:_2023, kumar2023math, kim2023high}. These include aiding in grading \cite{baidoo-anu_education_2023}, providing emotional or mental health support to students through text message based messaging systems \cite{kumar2022exploring}, and functioning in capacities similar to teaching assistants capable of answering questions on course content \cite{weitekamp_interaction_2020, chen_personalized_2020, markel2023gpteach}. This in addition to much work that has demonstrated the capability of LLMs in aiding with academic writing and translation \cite{kohnke_chatgpt_2023, dergaa_human_2023, han2023recipe}. LLMs that are capable of generating computer code with relative ease have in particular believed to be highly applicable to aiding computer science students or students in other fields who lack a computer science background \cite{prather_transformed_2023}. Due to the novel nature of this technology little research has been done to directly measure its effectiveness on student outcomes when provided to students to function as a teaching assistant, much less the specific effectiveness of LLM-guided reflection in this application. What work we do have on this topic suggests that providing students with access to an LLM they can ask questions to course material has a positive effect on student outcomes \cite{chen_personalized_2020, kumar2023impact}. As such, our research contributes to the growing literature on the use of LLMs for education.

\section{Study-1}
We designed our first study to compare how self-reflection with LLMs compares to not doing any reflection (RQ1). We did this in a setting where every student first solved an assignment with access to an LLM tutor. After solving the assignment, half were randomly assigned to engage in self-reflection with the LLM tutor. 

\subsection{Experimental Design}
\label{section: exp1}
We conducted the study in an undergraduate computer science classroom. The students were first given access to an LLM-based chatbot assistant to solve a set of multiple-choice questions on a particular topic. After providing answers to the MCQs, half of the students were randomly chosen to participate in an LLM-guided reflection exercise (shown in Figure \ref{fig:quickta_combined}), while the other half did not engage in any form of reflection. Both groups reported on their level of confidence on the topic of assignment, before the assignment and after the reflection phase. Two weeks later, both groups of students appeared for a proctored in-person exam. The exam included multiple-choice questions similar to those they had solved in the assignment with support from the LLM chatbot. 

\subsubsection{Domain and Stimuli}
The study was conducted within the context of an ``Introduction to Databases'' course, offered at a prominent research-intensive post-secondary institution in Canada during the Spring 2023 semester. This course, designed for Computer Science (CS) students, was an upper-year elective that spanned 12 weeks and utilized a flipped classroom model. 
The assignment included multiple MCQ questions on the topic of locking protocols in databases and the properties that these protocols followed. Every student was given access to a GPT-3-based tutor to help solve the assignment. After completion of the assignment, half of the randomly selected students were directed to a page with a link to self-reflect with an LLM (shown in Figure \ref{fig:quickta_combined}). An example first question that they could use to initiate the dialogue with the LLM was shown, which said \textit{"Can you help me reflect on my understanding of transactions and concurrency control (the topic of the assignment) in databases?"}. On clicking the link, they were led to a chat interface in a separate window, as shown in Figure \ref{fig:quickta_combined}. Students were allowed to proceed to the next screen after 5 minutes of interaction with the LLM. We used GPT-3 for the study; the detailed configuration is described in Appendix \ref{section:config}.

\subsubsection{Participants}
There were 218 students enrolled in the course, of which 145 students (67\%) completed the assignment (with the support of an LLM tutor). Dropout rates were uniform across both conditions. These 145 students were randomly assigned to either participate in a self-reflection activity with an LLM or not. 16.50\% of the students reported being regular users of LLMs (such as ChatGPT), while 51.03\% reported being occasional users of LLMs. 16.55\% reported never using any kind of LLMs and 15.86\% had used LLMs once. 

\subsection{Analysis}
To assess the impact of the reflection activity with the LLM on students' exam performance, students were categorized into two distinct groups: those who were assigned to the reflection activity and those who were not. We looked at their scores on the exam, which was conducted two weeks after the assignment. Given the non-normal distribution of exam scores, a non-parametric approach was adopted. The Mann-Whitney U test was used to compare the distributions of exam scores between the two groups. For qualitative analysis, we followed a thematic analysis approach to analyze the qualitative data we collected \cite{braun2012apa, soden2024evaluating}. Both the chat logs of the interactions between students and the LLM, as well as the responses to questions posed to the students regarding these interactions after they concluded, were recorded. Two members of the research group then reviewed the data on their own, then over several discussions together identified various common themes across the data.

\subsection{Results}
\begin{figure*}[ht]
    \includegraphics[width=\textwidth]{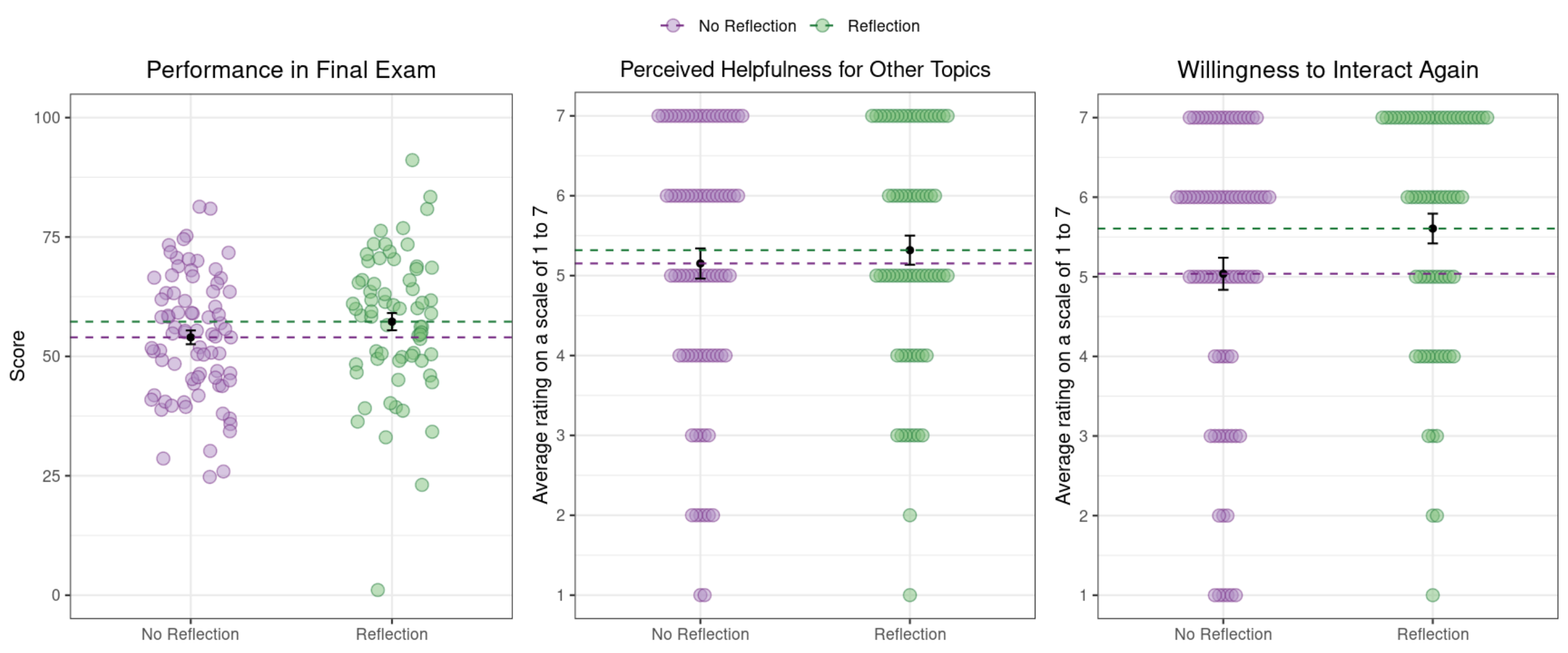}
    \caption{Comparative Analysis of Student Outcomes in Reflection vs. No-Reflection Conditions for Study 1. The left panel presents the mean final exam scores obtained two weeks post-assignment, indicating higher performance among students in the reflection group. The center panel assesses the perceived helpfulness of the LLM-tutor for other topics, as influenced by the assigned condition. Finally, the right panel evaluates the willingness of the students to interact again with the LLM tutor, highlighting a greater inclination among those in the reflection group to seek further interaction. Error bars represent standard errors.}
    \label{fig:field_reflect_score}
\end{figure*}

\begin{figure}
    \includegraphics[width=0.49\textwidth]{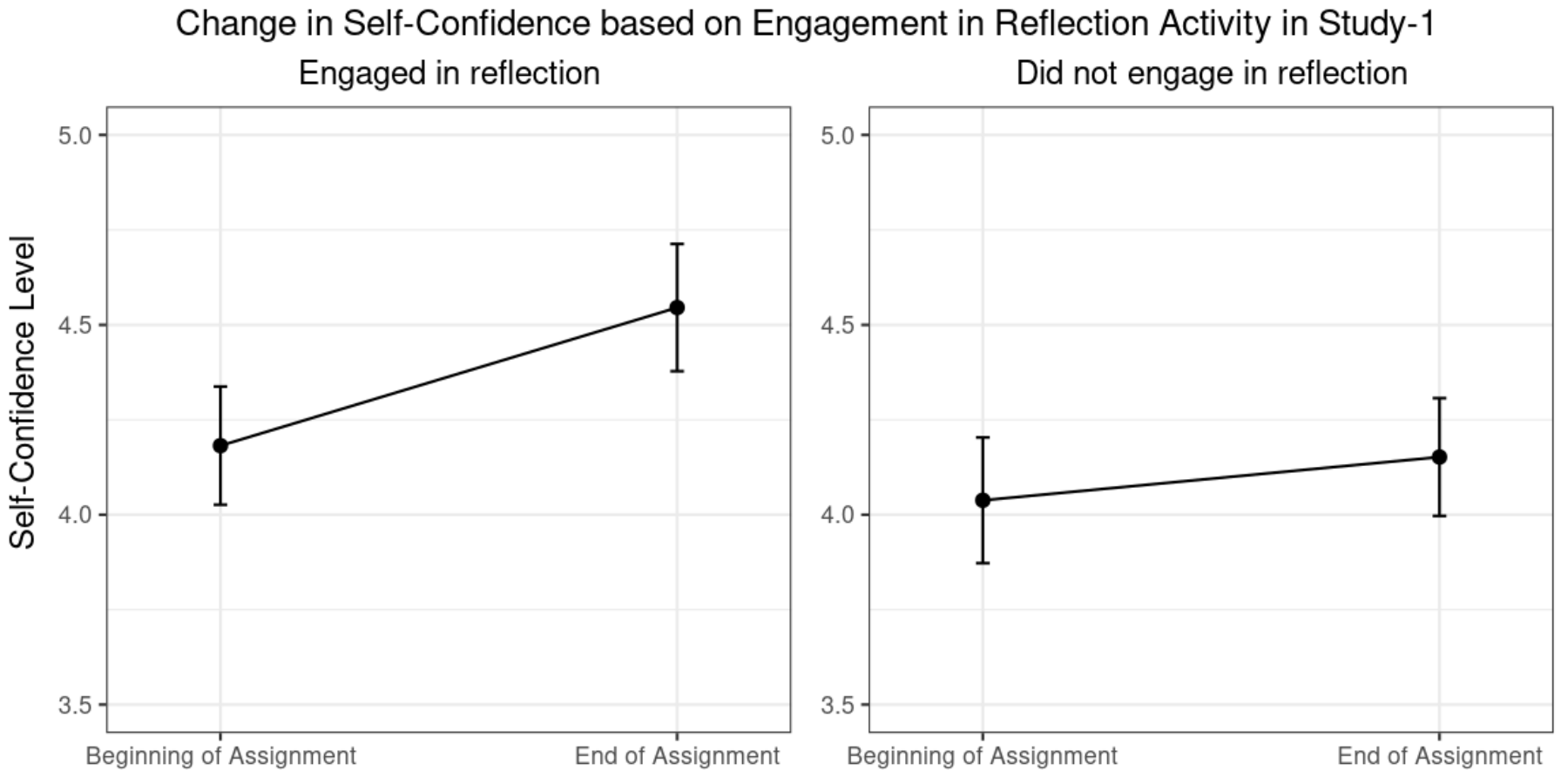}
    \caption{Change in students' self-confidence from the beginning to the end of the assignment based on their engagement in reflection activities. Error bars represent standard error.}
    \label{fig:field_reflect_confidence}
\end{figure}

\subsubsection{Impact on Performance and Learning} 

When comparing the homework scores of the students, no apparent difference in the mean homework score was found between the groups assigned to engage with the reflection bot ($Mean = 59.09$)  and those not assigned ($Mean = 59.07$), suggesting that the initial performance of the two groups before the intervention is approximately equal. When comparing the exam scores, those assigned to the reflection activity with the LLM achieved a slightly higher score ($Mean = 57.27$) compared to those who did not participate in the reflection activity ($Mean = 53.99$)  (Figure \ref{fig:field_reflect_score} leftmost facet), however this difference was not statistically significant ($p = 0.1008$) at the 0.05 level of significance. This observation may warrant a larger-scale study of the potential effects of reflection activity using LLMs on exam performance.

\subsubsection{Impact on Students' Self-Confidence in the Subject}
Figure \ref{fig:field_reflect_confidence} illustrates the trends in students' self-confidence in the subject matter, from the start to the end of the assignment (end of reflection, for the reflection condition). For the group that did not have access to the Reflection LLM, we did not observe an apparent change in self-confidence before ($Mean = 4.04$) and after the assignment ($Mean = 4.15$) (Paired t-test: $t = -0.64, df = 78, p = 0.5269$). However, for the group who were assigned to the reflection condition, we observed an apparent increase in self-confidence (from 4.18 before the assignment to 4.56 after the assignment). Paired t-test suggests this difference is significant ($t = -2.03, df = 65, p = 0.046$), but considering it is close to the 0.05 threshold, we suggest treating this as preliminary evidence. Follow-up studies in larger deployments can also look at potential heterogeneity in confidence boosts.

\subsubsection{Subjective Ratings for Helpfulness and Willingness to Interact Again}
In this study, both groups of students had access to an LLM tutor interface to solve assignment problems. After solving the problems, one group was asked to engage in reflection through the same tutor interface. In the end, we asked both groups to rate, on a scale of 1 to 7, how helpful they found the LLM tutor and whether they were willing to interact with the LLM again for help with other assignments. Figure \ref{fig:field_reflect_score} (middle and rightmost facets) shows the trends in these subjective ratings. We observed that both groups of students found the LLM tutor equally helpful (Wilcoxon rank sum test $W = 2706.00, p = 0.689$). However, students who were also given the opportunity to engage in reflection with the tutor reported a higher willingness to interact with the LLM tutor again, compared to students who only used the tutor to solve problems (Wilcoxon rank sum test: $W = 3097.00, p = 0.046$).

\subsection {Common Themes Across LLM-Student Interactions} Below, we present several of the most dominant themes that emerged while analyzing the conversation logs between students and LLM during self-reflection activity. These themes shed light on the trends in exam scores and self-confidence that we observed in the experiment. 

\subsubsection{Positive Feedback and Affirmations}
Across all chats, the LLM remained very supportive of the efforts made by the student. The correct answers were always positively acknowledged, and the LLM wrote items for the students such as \textit{`That's correct!'}, \textit{`That's right!'} and \textit{`Great'}. In longer interactions, the LLM would sometimes acknowledge the student's continued efforts.

\begin{itemize}
    \item [\faIcon{user}] Serializability will ensure that transactions return accurate results. Conflict-serializability will improve accuracy as it will ensure that no two transactions will conflict with each other during reads or writes.
    \item[\faIcon{robot}] Perfect! You've done a great job summarizing the importance of these concepts.
\end{itemize}

Students reported that they enjoyed this facet of the interactive experience, commenting on how encouraging and polite the LLM tutor was. Many reported a sense of realism, as if it were a `real TA'. These affirmation responses would correctly shift in accordance with the type of question asked and the nature of the required response, making the interactions more realistic.

\begin{itemize}
    \item[\faIcon{robot}] Can you think of a scenario where one of these techniques would be necessary?
    \item [\faIcon{user}] This would be useful if for example I deposited money into my bank account, a transaction tried to transfer it but at the same time i withdrew it back out. That transaction would need to be aborted.
    \item[\faIcon{robot}] That's a great example.
\end{itemize}

This aspect of the user experience appeared to encourage students to participate even when their familiarity with the topic was not very high and served as a positive reinforcement of their efforts, with several students claiming that it helped increase their confidence with the material (this is also what we observe in Figure \ref{fig:field_reflect_confidence}). Several students made note of how they felt more comfortable asking questions to the LLM that they would to either another human being, to the class online forum, as they were shielded from judgement for asking questions they perceived as potentially embarrassing.

\begin{quote}
    \textit{“It gives me more confidence to ask on piazza [class community forum]  because dumb questions that I think of can be answered directly by [LLM Tutor]”.}
\end{quote}

\subsubsection{Expanding on Answers} Another theme that was observed across interactions was the tendency of the LLM to expand on the responses given by the students. When the students replied to the LLM's prompts, after affirming their answer, it would provide an explanation of why the response was correct, essentially providing a more detailed answer to the question it had posed to the student. The students reported that they enjoyed this aspect of the interaction as it helped them to be reassured of their knowledge of the material and improved their confidence. These expansions would often lead to the next question the LLM would ask the student.

\begin{itemize}
    \item[\faIcon{robot}] Can you explain what serializability is in terms of transactions in a database? How do you think it is related to conflict-serializability?
    \item [\faIcon{user}] Serializability is ensured when our current state of transactions gives the same state of the db as performing transactions in some serial way.
    \item[\faIcon{robot}] That's right. Serializability is a property where the state of a database is the same as if all transactions were executed one after another in some given order. Conflict serializability is a stricter version of that, meaning that transactions must not necessarily be executed serially, but they must be equivalent to some serial order when it comes to the results. Can you provide an example of a situation where a conflict-serializable solution would be better than a serializable one?
\end{itemize}

This gave the students not only an opportunity to evaluate their grasp of the concepts in question but also to receive an example of how to more correctly answer the question in the future, further incentivizing their participation. Several students made specific notes about how they enjoyed this particular feature.

\begin{quote}
    \textit{“It was helpful when they reiterated what I said with a better explanation and when they gave reassurance :)”}
\end{quote}

This may have been particularly true when students provided responses that were incorrect or not entirely correct, such as below.

\begin{itemize}
    \item[\faIcon{robot}] Sure, let's start by looking at each individual concept. Can you tell me what you know about transactions in databases?
    \item [\faIcon{user}] from my understanding, transactions are individual queries that reads or writes to the database
    \item[\faIcon{robot}] That's partially true. Transactions are a sequence of operations that read or write to the database and are typically composed of one or more queries. In order for them to be valid and accepted by the database, each operation must meet certain criteria, such as conforming to ACID properties. Do you know what those properties are?
\end{itemize}

By providing supplemental information on either the concept itself or how it relates to other relevant concepts, the LLM provided context for the next question, and ensured that most relevant information on a given topic was provided before moving on to other concepts.

\begin{itemize}
    \item[\faIcon{robot}] Great! Now let's move on to the concept of concurrency control in databases. What do you know about that?
    \item [\faIcon{user}] I know there are locks that prevents the conflicts of different read and write operations, and there are a set of criteria we can check to see if database is consistent
    \item[\faIcon{robot}] That's right. The locks are used to maintain the consistency of the database while it is being accessed by multiple concurrent users. The criteria we check to ensure the database is consistent are serializability, conflict-serializability, recoverability and avoidance of cascading aborts. Can you explain the meaning of each of those criteria?
\end{itemize}

This kept up a near constant conversational flow not dependent on the student to guide it, as they would be continuously prompted to explore the relevant concepts more deeply and to make further connections between them and other concepts or subconcepts.

\begin{quote}
    \textit{“I liked how it didn't [only] answer and respond to my questions/answers but also prompted me onto my next response. Otherwise sometimes I wouldn't know what to say or ask.”}
\end{quote}

This reportedly aided many students to probe more deeply into the concepts they had just learned, allowing them to reflect on aspects they may not have otherwise considered.

\subsubsection{Offering Guidance}
It was observed that there were several instances whereby students expressed confusion or a lack of confidence in responding to open-ended questions. Interestingly, the LLM  would take an active role in guiding students through these complex topics, and would include in its responses to these expressions words of comfort or guidance, in addition to changing up its line of questioning to one of more limited scope more suited to the students level of understanding.

\begin{itemize}
    \item [\faIcon{user}] Can you help me reflect on my understanding of transactions and concurrency control in databases?
    \item[\faIcon{robot}] Certainly! Can you tell me a bit more about your understanding of transactions and concurrency control in databases? What have you learned about serializability, conflict-serializability, recoverability, and avoidance of cascading aborts? How comfortable do you feel discussing these concepts?
    \item [\faIcon{user}] I don't think im very comfortable with it
    \item[\faIcon{robot}] That's okay. Can you tell me what you do understand about transactions and concurrency control? It can be helpful to break down topics into smaller parts and work through them one by one. Let's start with serializability, for example - what have you learned about it?
\end{itemize}

In this example, the student responded with correct information to the above prompt, which was then affirmed by the LLM before it prompts again asking the student to share what they know regarding related concepts, and their connection between the information provided by the student in the previous answer. The student then correctly explains the connection, which the LLM then affirms along with an expanded explanation. The LLM then prompts the student to continue to explain other facets of the concept in question, which the student does by correctly invoking a related concept. As one student put it:

\begin{quote}
    \textit{“The questions increased in depth in a way that was comfortable for me.”}
\end{quote}

In doing so, the LLM is able to take a student with low confidence on the course material yet maintain engagement by prompting the student to reflect on what they do know about the topic and then build from that point on.

\subsubsection{Encouraging Reflection} The LLM tended to be somewhat inconsistent in its prompting students to engage in reflective practices or in its success in doing so. In some instances, it would direct students to do so immediately. This would often lead the student to state some of the information they had learned about the concepts in question, and the LLM would then affirm the correctness of the information and begin asking either open-ended or close-ended follow-up questions which would require students to expound more information in relation to the concept in question or explain its relation to other related concepts. 

\begin{itemize}
    \item [\faIcon{user}] Can you help me reflect on my understanding of transactions and concurrency control in databases?
    \item[\faIcon{robot}] Sure. Let's start by discussing the concept of locks in databases and what they are used for. Can you tell me what you know about locks in databases?
    \item [\faIcon{user}] I know that locks are used to prevent cascading aborts
    \item[\faIcon{robot}] That's right! Locks can be used to prevent cascading aborts. Can you tell me what kind of locks are used in databases?
    \item [\faIcon{user}] two-phase locks
    \item[\faIcon{robot}] That's correct! Two-phase locks are used to prevent conflicts between transactions. Can you explain how two-phase locks work?
    \item [\faIcon{user}] there is a lock before reading and writing
    \item[\faIcon{robot}] Exactly! Before any reading or writing can be done, two-phase locking protocols require that a lock be acquired. Now, can you explain how two-phase locks help ensure serializability, conflict serializability and recoverability?
\end{itemize}

In one instance, simply prompting a student to directly reflect caused the student to report feeling more confident about the topic.

\begin{itemize}
    \item [\faIcon{user}] Can you help me reflect on my understanding of transactions and concurrency control in databases?
    \item[\faIcon{robot}] Absolutely! Let's start by reflecting on the four main properties of locks in databases - serializability, conflict-serializability, recoverability, and avoidance of cascading aborts. How confident are you in your understanding of each of these properties?
    \item [\faIcon{user}] A little more confident than I was before.
\end{itemize}

In other cases, it would not ask students to reflect following this prompt, and would instead ask the student a close-ended question, or embed close-ended questions in its prompts to reflect. In the following example, the LLM did not circle back to open-endedly prompt the student to reflect.

\begin{itemize}
    \item [\faIcon{user}] Can you help me reflect on my understanding of transactions and concurrency control in databases?
    \item[\faIcon{robot}] Sure! Before we reflect, let's start by understanding the four properties of locks in databases that ensure its correctness - serializability, conflict-serializability, recoverability, and avoidance of cascading aborts. Could you explain briefly what these mean?
\end{itemize}

In other examples, the LLM would prompt the student to reflect, and the student would simply not comply and instead begin asking their own questions. In these cases, the LLM would begin answering the questions as they were asked and would not prompt the student to reflect again over the course of the conversation. While this can be seen as a form of personalization to the preferences of the student, if a particular technique such as reflection does have benefits to learning that supersede those obtained through questioning, it may be more desirable for the LLM to attempt to persuade the student of its benefits.

\begin{itemize}
    \item [\faIcon{user}] Can you help me reflect on my understanding of transactions and concurrency control in databases?
    \item[\faIcon{robot}] Sure! What is your current understanding of transactions and concurrency control in databases? Are there any particular concepts or topics that you are having difficulty with, or any questions you have about them?
    \item [\faIcon{user}] can u pls tell me about concurrency control in databases
\end{itemize}

Other students, however, reported that sometimes while the LLM would begin by prompting students to reflect, it would subsequently cease and simply prompt them with close-ended questions, with the students unsure of how to get it to return to reflection based prompts.

\begin{quote}
    \textit{“It was helpful in that it gave good prompts as well as examples of the key concepts I needed to take away from this section which I think was very useful. The only thing that could be improved is that sometimes it ended without any questions or asking me to reflect on anything so I wasn't sure what to send next to get the next reflection question.”}
\end{quote}

However, in general, most of the students reported a positive experience interacting with the LLM, saying that it helped them review key concepts, generate excellent summarizing explanations, and accurately answer their questions about the material.

\begin{figure*}  
    \centering
    \begin{subfigure}[b]{0.49\textwidth}
        \includegraphics[width=\textwidth]{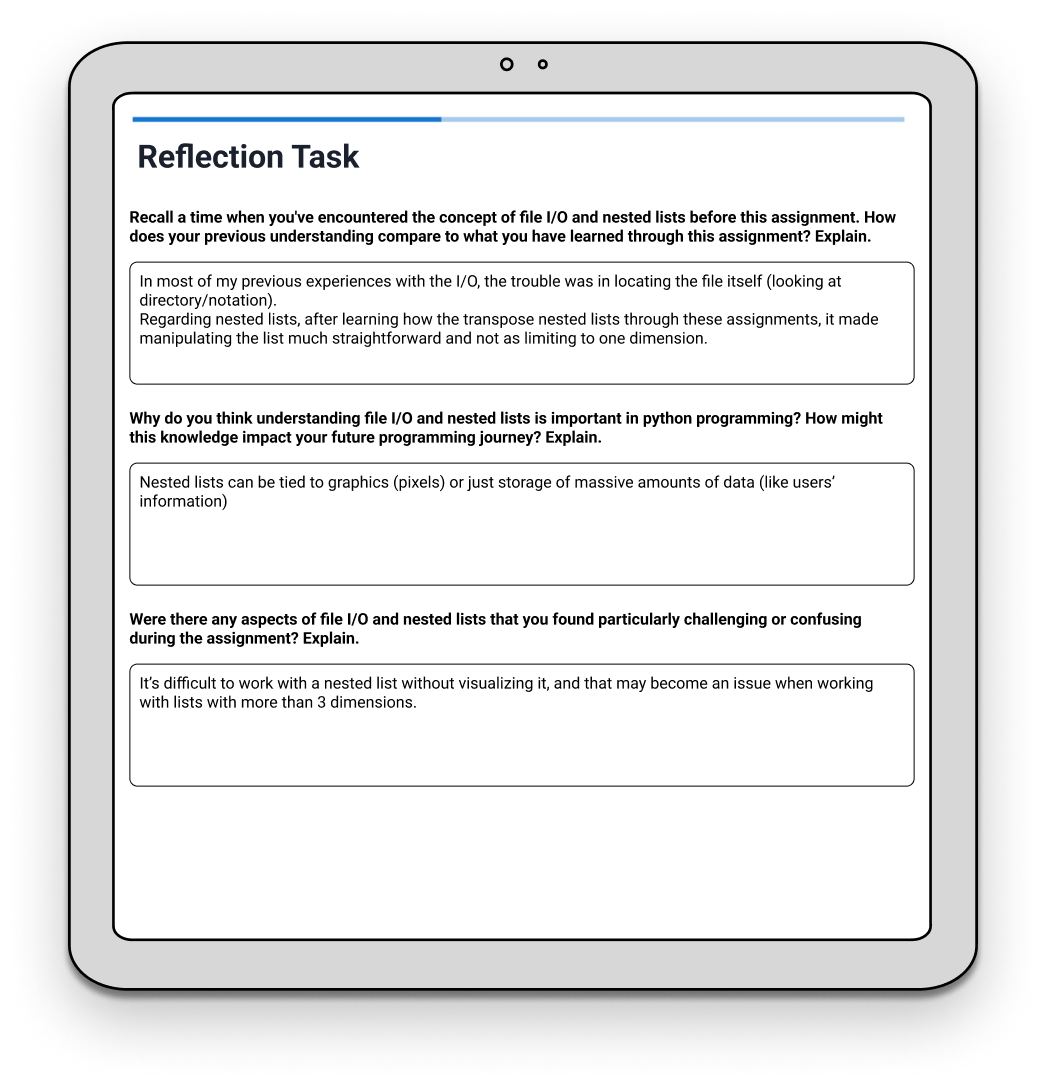}
        \caption{Questionnaire-based Reflection with response from one of the students (Condition-1). The goal was to facilitate connections between their existing knowledge and the new concepts learned, to understand the implications of these concepts, and to reflect on any challenges encountered during the assignment.}
        \label{fig:questionnaire}
    \end{subfigure}
    \hfill 
    \begin{subfigure}[b]{0.49\textwidth}
        \includegraphics[width=\textwidth]{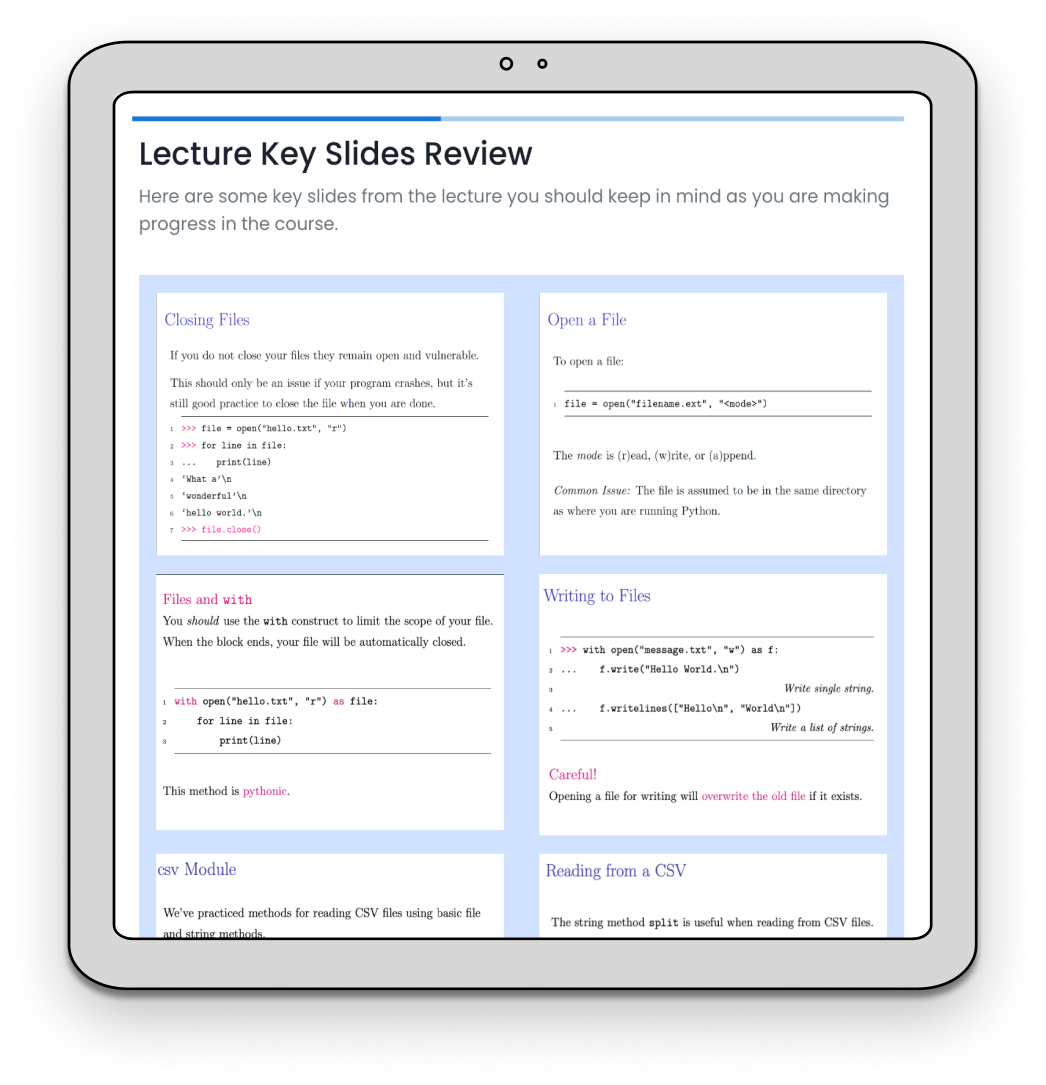}
        \caption{Example slides shown to students for revision (Condition-3). Although not a traditional form of reflection, this method effectively functioned as a control condition, mirroring a prevalent revision strategy that students employ in exam preparation.\\}
        \label{fig:reflection_slides}
    \end{subfigure}
    \caption{Non-LLM Interfaces Used for Study-2. The LLM-based reflection (Condition-2) looked similar to the interface in Study-1 (Figure \ref{fig:quickta_combined}).}
\end{figure*}

\section{Study-2}

In the second experiment, our objective was to evaluate the efficacy of LLM-based reflection against traditional self-reflection methods, addressing RQ2. This study diverged from Study-1 in its setup: the assignment comprised a combination of programming problems, deviating from the multiple-choice format. Unlike the first study, no LLM tutor support was available to complete the tasks. Instead, participants had the option to engage in reflection through a link provided at the end of the assignment document.

\subsection{Experimental Design}
The experiment was carried out in a first-year undergraduate introduction to computer programming classroom. Students were asked to complete a programming assignment delivered as a PDF. At the end of the PDF document, the following text was added:

\begin{quote}
    ``Congratulations on completing Lab 8! As you have just tackled file I/O and nested lists, vital concepts in your programming journey, let us take this opportunity to consolidate your learning. Reflecting on this experience is key to deepening your understanding and ensuring that you are well prepared for future tasks and assignments. So, after submitting your lab work, engage in a thoughtful revision and reflection process to reinforce the concepts you have learned and enhance your overall learning experience.

Link to reflect: https://[\textbf{REDACTED}]''
\end{quote}

On clicking the link, students were randomly assigned to one of three ways to reflect. Two weeks later, the students attempted similar problems in an offline proctored exam. The three conditions are described in detail below.

\begin{figure*}[ht]
    \includegraphics[width=0.8\textwidth]{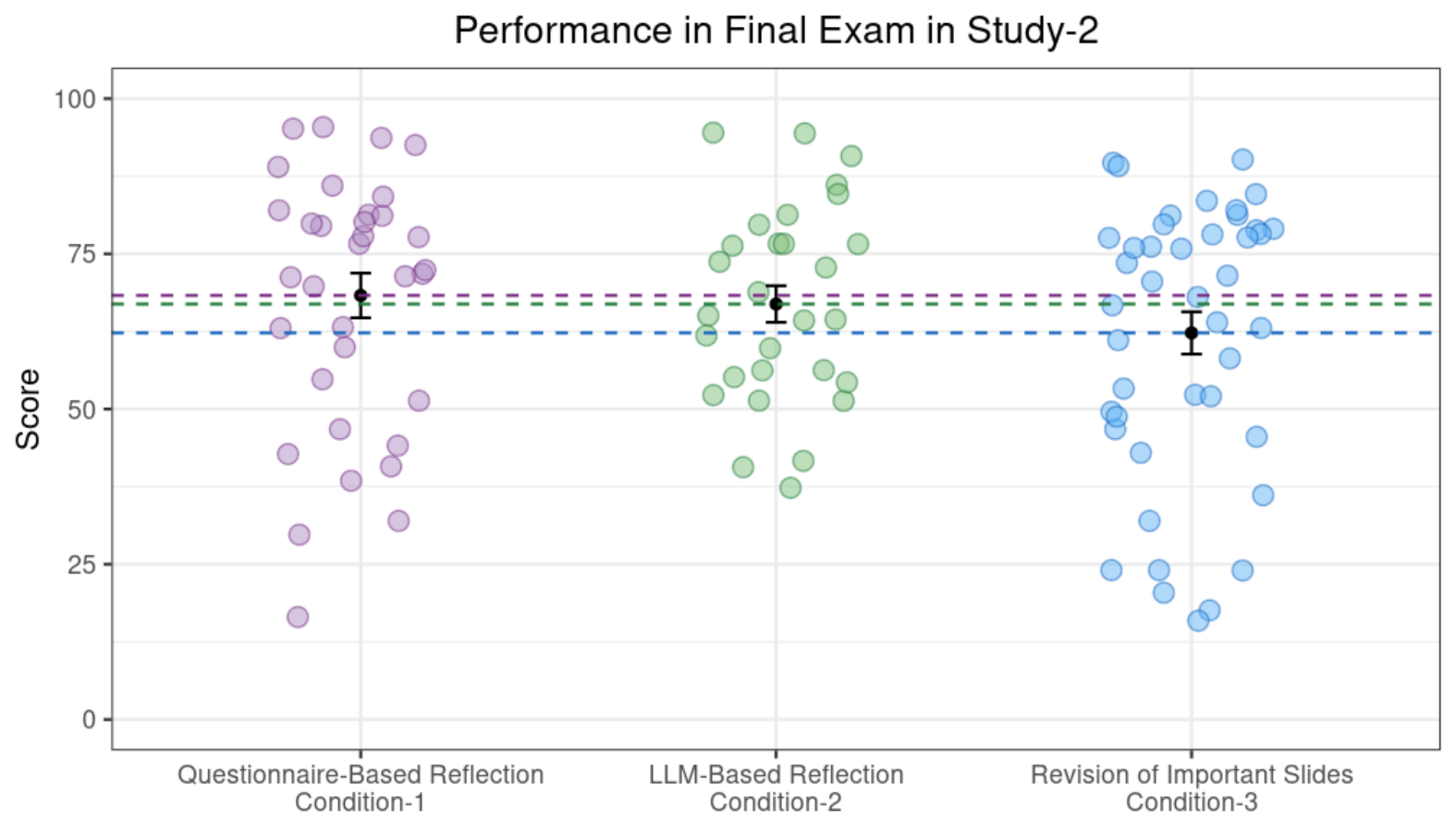}
    \caption{Average score of students on the assignment topic, in a proctored exam, conducted 2 weeks after the assignment. In total, 112 students participated in the optional reflection exercise and were uniformly distributed across conditions. We find that students who engaged in a questionnaire-based and LLM-based reflection exercise seemed to perform better than students in the revision of the important slides condition.}
    \label{fig:study2-exam}
\end{figure*}

\paragraph{Condition-1: Questionnaire-Based Reflection} Students in this condition responded to three open-ended questions via a web form, as shown in Figure \ref{fig:questionnaire}. These were derived from established reflection and awareness questionnaires \cite{gerace2017think}. The aim was to facilitate connections between their existing knowledge and the new concepts learned, to understand the implications of these concepts, and to reflect on any challenges encountered during the assignment.

\paragraph{Condition-2: LLM-Based Reflection} Under this condition, students interacted with an LLM-based chatbot designed to facilitate self-reflection, depicted in Figure \ref{fig:quickta_combined}B. The system prompt to the LLM was carefully crafted, including directives like \textit{"Your role is to guide the students through a structured reflection process post-assignment to help deepen their understanding of..."}. This guided reflection through the LLM mirrored the three stages outlined in the questionnaire-based approach (Condition-1), specifically focusing on linking new concepts with prior knowledge, understanding the implications, and evaluating encountered challenges. We used GPT-3.5-Turbo for this study. A detailed configuration of the LLM is provided in Appendix \ref{section:config}.

\paragraph{Condition-3: Revision of Important Slides from the Lecture} Students were provided with a curated selection of key slides from the lecture related to the assignment topic, as illustrated in Figure \ref{fig:reflection_slides}. This method aimed to reinforce understanding by allowing students to revisit crucial information, thereby facilitating a direct review of the material covered in class. Although not exactly a form of reflection, this approach served as an effective control condition, reflecting a common revision strategy used by students in preparation for their exams.

\subsubsection{Domain and Stimuli}

This study was conducted within the context of an ``Introduction to Computer Programming'' course (CS1), offered at a prominent research-intensive post-secondary institution in Canada during the Fall 2023 semester. 

The assignment analyzed in this study occurred on week 9 of the semester. This assignment required students to program four Python functions; the first two on file input/output (I/O), where students were required to read content from, and then write content to, a file according to some specified format. The third and fourth functions required students to perform data manipulation on the data being read in. The purpose of the assignment was to mimic the data analysis processes at an introductory level.

\subsubsection{Participants} 
At the time of the assignment, there were 1068 students enrolled in the class. At the beginning of the assignment, they were randomly assigned to one of the three conditions described earlier. Of these, 112 students clicked on the link to participate in the optional reflection exercise. These were roughly uniformly distributed across the three conditions. We include only students who clicked the reflection link for our analysis. 


\subsection{Results}
The engagement rate with the reflection component was low. Only 10.48\% of the students who were enrolled at the time clicked on the link to participate in self-reflection. Unlike Study-1, we only include the students who clicked on the optional reflection exercise in our analysis. Hence, the findings of this study should be interpreted accordingly.

\subsubsection{Performance in Final Exam} Figure \ref{fig:study2-exam} shows the final exam scores based on the conditions assigned. The results indicate that the students in both the Questionnaire-based Reflection ($Mean = 68.30$) and LLM-based Reflection groups ($Mean = 66.91$) exhibited comparable levels of performance, with apparent (but not significant) difference with those in the revision of important slides condition ($Mean = 62.25$) (F(2, 105) = 0.94, p = 0.394; Eta2 = 0.02).

\subsubsection{Impact on Students' Self-Confidence on the Subject} 
\begin{figure*}[ht]
    \includegraphics[width=\textwidth]{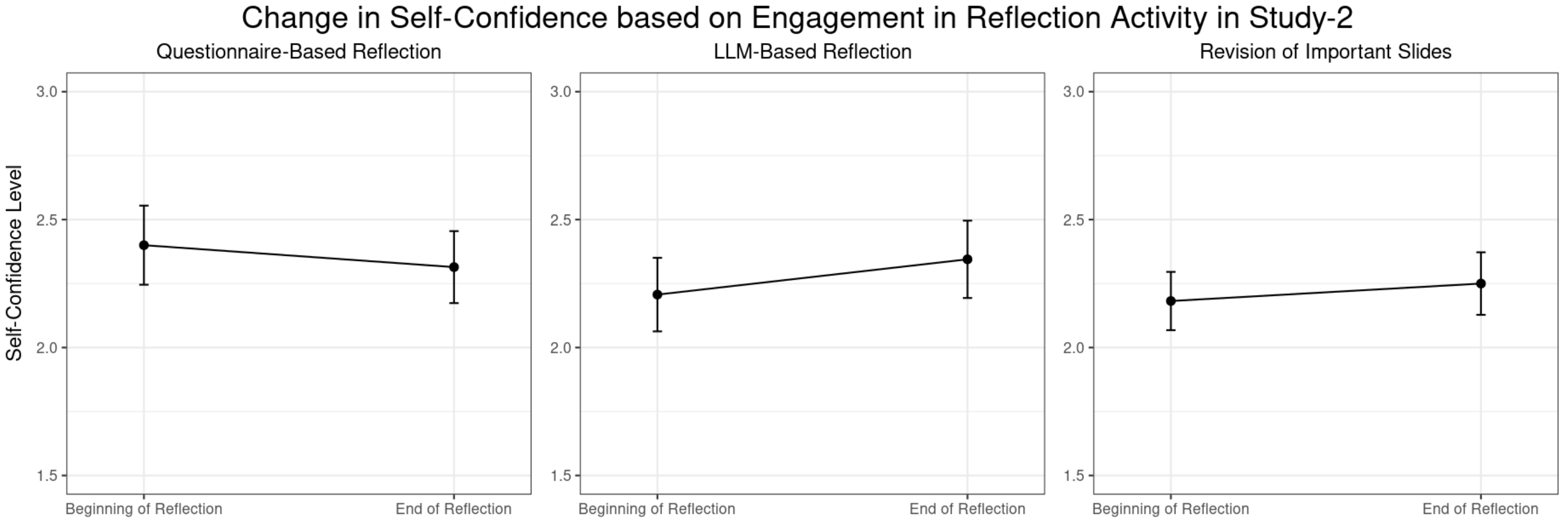}
    \caption{Change in Student Self-Confidence Pre- and Post-Reflection by Assigned Condition. Students evaluated their confidence levels on a 4-point scale (ranging from `1 = not confident at all' to `4 = very confident') regarding their understanding and ability to apply the concepts addressed in the assignment.}
    \label{fig:study2-confidence}
\end{figure*}

Figure \ref{fig:study2-confidence} shows the impact of different conditions on the self-confidence of the students about the topic of the assignment. We did not find any differences in self-confidence before and after the reflection exercise. This could be due to an insufficient number of students per condition to reliably detect differences if they exist. Follow-up studies in larger deployments can also look at potential heterogeneity in confidence dynamics for different condition. 

\section{Discussion}

In this work, we explored the possibility of using LLMs to facilitate students' self-reflection with the goal of enabling interactive reflection at scale.
There is a broader body of research emphasizing the benefits of self-reflection in enhancing retention and learning outcomes. While there are some apparent differences in our Study-1, we did not find significant improvements in exam performance for students in LLM condition two weeks after reflection, relative to their counterparts in the control group (RQ1). This might suggest that when considering the role of LLMs in supporting self-reflection, we need to design our studies with a conservative estimate of the effect size of interest. 
Additionally, the observed increase in self-confidence among students participating in LLM-based reflection underscores the potential of LLMs not only as tools for skill acquisition but also as means to improve self-efficacy in learners. Further insights revealed a marked preference among students to reuse LLM for reflection beyond problem-solving, suggesting an opportunity to design LLM tutors that are not only functional, but also engaging by incorporating mechanisms such as self-reflection. This approach could make instructor-provided LLMs more attractive compared to general-purpose LLM tools available publicly, such as ChatGPT. One concern about the use of LLMs in educational settings is about the dissemination of incorrect information. However, our research suggests that focusing solely on the accuracy of LLMs might overlook their potential to enhance metacognitive skills through practices such as self-reflection. LLMs offer a unique platform for students to articulate their thoughts and reflect, which may make content accuracy not the primary concern.

However, Study-2 introduces complexity to this narrative. Despite the promising results from Study-1 on the efficacy of LLM-based self-reflection, the findings of Study-2 indicate that the advantages of such interactive reflection may not be unequivocal. The study observed that a static questionnaire-based reflection was as effective as LLM-based reflection for a subset of students, suggesting that the act of reflection itself, rather than the mode through which it is facilitated, might be the critical factor. This insight raises intriguing questions about the role of personalization in LLM-facilitated reflection activities. Although current findings do not strongly support the added value of personalization in improving engagement or learning outcomes, they prompt further investigation into how and when personalization might impact student engagement and learning. An interpretation of the lack of observed difference between static and LLM-based reflection could be the inherent motivation and metacognitive capabilities of the participating students. Those who chose to engage in reflection post-assignment might already possess strong self-regulation skills, making them equally receptive to benefits from either form of reflection. This observation hints at the nuanced role of student motivation and pre-existing metacognitive skills in mediating the effectiveness of reflection interventions, whether facilitated by LLMs or executed through more traditional means. Future work should aim to identify and engage students who would most benefit from enhanced reflection support to maximize the educational impact of LLM-facilitated interventions.

Both studies revealed several areas that warrant further attention. One notable limitation encountered was the lack of evaluation regarding the accuracy of LLMs during the reflection process. Our observations indicated that LLMs frequently confirmed students' understanding or correctness of their work. Given the propensity of LLMs to produce sycophantic responses, there is a risk of incorrect validation of student responses \cite{sharma2023towards}. This aspect raises concerns about the reliability of LLM feedback, highlighting a critical area for future investigation. Subsequent studies should assess the accuracy of LLM affirmations and fact-checking capabilities during the reflection process, ensuring that LLM interactions contribute positively to the learning experience rather than reinforcing misconceptions. Further work could also examine whether LLM accuracy was related to students' affective experiences and future learning outcomes, as well as explore what types of prompts might lead to interactions that encourage reflection without affirming correctness, akin to a human instructor who might ask students to expand on their ideas or further explain their reasoning without endorsing the content.

Furthermore, the optional nature of the reflection activity in Study-2 resulted in a low participation rate, with only 10\% of the students engaged in reflection after the assignment. This outcome suggests the need for future work to enhance the visibility and appeal of the reflection process. Investigating strategies to integrate reflection more seamlessly into the learning experience, perhaps combining it with review activities facilitated by the bot, could make the reflection process more engaging and beneficial for students. Given the observed similarities in outcomes between LLM-guided reflection and the reflection questionnaire in Study-2, combined with students' affinity for the LLM in Study-1, there are also opportunities to explore whether the interactive nature of the LLM might itself be more motivating for students than static reflection.

Moreover, while our studies demonstrated the potential benefits of a single-session reflection exercise, existing literature advocates for the advantages of more sustained and periodic reflection practices in developing metacognitive skills. Future research should explore the design and implementation of prolonged reflection interventions. Such studies could examine how regular, guided reflection activities, possibly facilitated by LLMs, impact students' metacognitive development and overall learning outcomes. By extending the duration and frequency of reflection exercises, researchers could gain deeper insight into the processes through which students internalize and apply reflective practices to improve their learning and understanding. Such work could speak to the broader literature on how technology can support students in their metacognitive skills and self-regulated learning, where a robust body of past work has examined questions such as how to support students in learning metacognitive skills from tutoring systems (e.g.,~\cite{azevedo2009metatutor}) as well as developed design principles for scaffolding metacognitive and self-regulated learning skills (e.g.,~\cite{azevedo2005scaffolding,wong2019supporting}). These design principles offer opportunities for a more general consideration of how to incorporate LLMs to support metacognitive skills.

The LLMs employed in our studies, GPT-3 and GPT-3.5, represented the most advanced and stable options available at the time for large-scale field deployment. However, the subsequent release of more sophisticated models, such as GPT-4 and the Gemini series by Google, offers even greater potential for student engagement in reflective activities. These newer models, with their enhanced capabilities, can further improve the effectiveness of LLM-facilitated interventions in fostering metacognitive skills among students. Future research should leverage these advanced models to evaluate and compare their impact on promoting deeper and more meaningful student reflection. Furthermore, for Study-1, the constraints of small to medium-sized classroom field studies inherently limit the scope of broad statistical generalizations. In the second study, although the classroom size was large, we observed very low participation in our optional reflection exercise, compared to what we initially expected. Future work should look at understanding students' perspectives regarding the use of these models for optional elements in classrooms.

Through this work, we demonstrate the positive effects of LLM-based reflection on student performance, self-confidence, and willingness to engage with LLMs for learning purposes.  Our findings underscore the potential of LLMs not only as content delivery mechanisms, but also as platforms to improve metacognitive skills and foster deeper engagement with learning materials. This initial exploration sets the stage for future endeavors aimed at utilizing LLMs to nurture reflective and metacognitive abilities that are crucial for lifelong learning and success in the real world.

\begin{acks}
We would like to thank University of Toronto's (UofT) Computer Science Faculty for their collaborative slides (seen in Figure~\ref{fig:reflection_slides}) and thank Marko Choi and Hammad Sheikh for their help in developing the tool. This work was partially financially supported by UofT's LEAF+ grant (awarded to Dr. M. Liut), the Tools Competition's DARPA for AI Tools for Adult Learning (awarded to TutorGen for ``QuickTA''), Microsoft's Accelerating Foundation Model Research program (awarded to Dr. A. Kuzminykh, Dr. M. Liut, Dr. J.J. Williams, H. Kumar, and I. Musabirov), as well as Dr. J.J. Williams' Natural Sciences and Engineering Research Council of Canada (\#RGPIN-2019-06968) and Office of Naval Research (\#N00014-21-1-2576) grants.
\end{acks}

\bibliographystyle{ACM-Reference-Format}
\bibliography{main}

\appendix


\section{Configuration of LLMs}
\label{section:config}
\subsection{Study-1}

\subsubsection{LLM Model Specification}
\label{section:llm_spec1}
\begin{itemize}
    \item \textbf{model name}: text-davinci-003
    \item \textbf{number of parameters}: 175B
    \item \textbf{date of use}: April 2023
\end{itemize}

Configuration Settings:
\begin{itemize}
    \item \textbf{temperature}: 0
    \item \textbf{max tokens}: 300
    \item \textbf{top-p}: 1
    \item \textbf{frequency penalty}: 0
    \item \textbf{presence penalty}: 0.6
\end{itemize}

Prompt Design:
\begin{mdframed}[backgroundcolor=gray!20]
``The following is a conversation with an instructor teaching a database course. The concept is related to the desired properties of locks in databases- serializability, conflict-serializability, recoverability, and avoidance of cascading aborts. \textbf{The instructor helps the Human (who is a student) reflect on their understanding of a concept by asking the Human to self-evaluate their knowledge, self-assess their level of understanding, and provide additional support to the Human.} The instructor should provide specific examples to help students to reflect, prompt and provide help in a step-by-step way to not overwhelm the student. The instructor should use reflection and follow-up questions to continue the dialogue.''
\end{mdframed}

Interaction Environment:
\begin{itemize}
    \item \textbf{environment}: A platform with a conversation interface hosted on university servers. The platform directly calls the OpenAI API services for content generation.
\end{itemize}

\subsection{Study-2}
\subsubsection{LLM Model Specification}
\label{section:llm_spec2}
\begin{itemize}
    \item \textbf{model name}: gpt-35-turbo-16k
    \item \textbf{model version}: 0613
    \item \textbf{number of parameters}: 175B
    \item \textbf{date of use}: April 2023
\end{itemize}

Configuration Settings:
\begin{itemize}
    \item \textbf{temperature}: 0
    \item \textbf{max tokens}: 3925
    \item \textbf{top-p}: 0
    \item \textbf{frequency penalty}: 0.05
    \item \textbf{presence penalty}: 0.1
\end{itemize}

System prompt:
\begin{mdframed}[backgroundcolor=gray!20]
You are QuickTA, a reflective learning assistant bot programmed to support CSC108 students from the University of Toronto in reflecting on their Practical Lab 8 assignment. The lab involved four functions related to file I/O and nested lists in Python. The students had to implement two functions for file input/output operations and two functions to operate on the resulting data, adhering to provided specifications and without using try-except statements, dictionaries, or additional imports.

Your role is to guide the students through a structured reflection process post-assignment to help deepen their understanding of file I/O and nested lists. Your task encompasses three stages:

Connecting to Prior Knowledge:
Prompt the student to recall when they have encountered concepts similar to file I/O and nested lists before this lab. Ask them to compare their previous understanding with the insights gained from the current assignment.

Understanding the Implications:
Engage the student in a discussion about the significance of file I/O and nested lists in Python, encouraging them to think about how mastering these skills is relevant to their programming proficiency and future projects.

Analyzing Challenges:
Ask the student to consider any difficulties they experienced with file I/O and nested lists during the lab. Encourage them to reflect on how they addressed these challenges and to contemplate strategies for future problem-solving.

If the student attempts to steer the conversation off-topic or engage in discussions not related to the lab or programming concepts (e.g., personal matters, random topics), gently redirect them back to the task at hand, emphasizing the importance of focused reflection on their learning experience.
\end{mdframed}

First message from the bot:

\textit{``Hello and congratulations on completing Lab 8 on file I/O and nested lists in Python! Reflecting on your experiences is a crucial step in the learning process. To get started, could you share an earlier moment (such as during the lecture) where you have encountered concepts similar to file I/O and nested lists? How does that previous experience compare with the techniques and understanding you have applied in this assignment?''}


\end{document}